\documentclass[aps,twocolumn,pre,floatfix]{revtex4-2}

\usepackage{xcolor,hyperref}
\hypersetup{
   colorlinks,
   linkcolor={blue!50!black},
   citecolor={blue!50!black},
   urlcolor={blue!80!black}
}

\usepackage{epsfig, amsmath}
\usepackage{bm}
\usepackage{soul,ulem}
\usepackage{hyperref}
\usepackage{footmisc}
\usepackage{graphicx,enumitem}
\DeclareMathAlphabet{\mathitbf}{OML}{cmm}{b}{it}

\renewcommand{\=}{\!=\!}

\newcommand{\Psiv}{\bm{\Psi}}

\begin{document}

\title{Testing the Heterogeneous-Elasticity Theory for \\ low-energy excitations in structural glasses}
\author{Edan Lerner$^{1}$}
\author{Eran Bouchbinder$^{2}$}
\affiliation{$^{1}$Institute of Theoretical Physics, University of Amsterdam, Science Park 904, 1098 XH Amsterdam, the Netherlands\\
$^{2}$Chemical and Biological Physics Department, Weizmann Institute of Science, Rehovot 7610001, Israel}

\begin{abstract}
Understanding the statistical mechanics of low-energy excitations in structural glasses has been the focus of extensive research efforts in the past decades due to their key roles in determining the low-temperature mechanical and transport properties of these intrinsically nonequilibrium materials. While it is established that glasses feature low-energy nonphononic excitations that follow a nonDebye vibrational density of states, we currently lack a well-founded theory of these fundamental objects and their vibrational spectra. A recent theory --- that builds on the so-called Heterogeneous-Elasticity Theory (HET) and its extensions --- provides explicit predictions for the scaling of the low-frequency tail of the nonphononic spectrum of glasses, the localization properties of the vibrational modes that populate this tail, and its connections to glass formation history and to the form of the distribution of weak microscopic (interatomic) stresses. Here, we employ computer models of structural glasses to quantitatively test these predictions. Our findings do not support the HET's predictions regarding the nature and statistics of low-energy excitations in glasses, highlighting the need for additional theoretical developments.
\end{abstract}

\maketitle

Following advances initiated about four decades ago (e.g.,~\cite{soft_potential_model_1987,karpov1983}), it is now well established that structural glasses feature quasilocalized, nonphononic low-energy (soft) excitations~\cite{JCP_Perspective}. These are manifested at low frequencies $\omega$ in the vibrational density of states (VDoS) ${\cal D}(\omega)$ of glasses, once hybridization with phonons --- that follow Debye's VDoS ${\cal D}_{\rm D}(\omega)$ --- is properly accounted for~\cite{SciPost2016,phonon_widths,enumarating_arXiv_2024,experimental_evidence_2024_arXiv}, taking the form ${\cal D}(\omega)\!\sim\!\omega^\beta$ for $\omega\!\to\!0$. The nonphononic excitations are intrinsically related to the disordered, noneqilibrium structures exhibited by glasses, e.g., dependent on their formation history. The nature and statistical properties of these nonphononic excitations are currently extensively investigated~\cite{JCP_Perspective,modes_prl_2016,ikeda_pnas,lerner2019finite,tanaka_2d_modes_2022,jcp_2d_vdos_2022,ikeda_2d_vdos_jcp_2023,smarajit_droplets,experimental_evidence_2024_arXiv,boson_peak_2d_jcp_2023,moriel2024boson}. A large body of evidence suggests that $\beta\=4$, independently of glass composition, interatomic interactions, thermal history, space dimensionality, and long-range order~\cite{modes_prl_2016,modes_prl_2018,modes_prl_2020,LB_modes_2019,lerner2019finite,Atsushi_high_D_packings_pre_2020,jcp_2d_vdos_2022,disordered_crystals_spectrum_2022,ikeda_pnas,universal_VDoS_ip,ikeda_2d_vdos_jcp_2023,Ning_2024_finite_size_nat_comm,experimental_evidence_2024_arXiv}.

As of now, there exists no well-established theory of these observations. Early attempts, collectively known as the ``Soft Potential Model''~\cite{karpov1982,klinger1983,soft_potential_model_1989,soft_potential_model_1991,buchenau_prb_1992,buchenau_Phil_Mag_1992,soft_potential_model_prb_1993,Chalker2003,KLINGER2010111}, tried to account for the nonphononic $\omega^4$ VDoS through the properties of a certain class of random functions that mimic the intrinsically disordered potential energy of glasses. Subsequent developments envisioned quasilocalized, nonphononic excitations as anharmonic oscillators that are embedded inside a 3D elastic continuum and that interact through dipolar forces with random amplitudes and orientations~\cite{Gurevich2003,Gurevich2005,Gurevich2007}, leading to a nonphononic $\omega^4$ VDoS. A recent reincarnation of these ideas led to the formulation of a mean-field model (no spatial embedding), where the anharmonic oscillators randomly interact among themselves and with a force field~\cite{scipost_mean_field_qles_2021,meanfield_qle_pierfrancesco_prb_2021,folena2022marginal,franz2022linear,urbani2022field,urbani2024statistical,maimbourg2024two,moriel2024boson}, also leading to a nonphononic $\omega^4$ VDoS.

A different approach, known as the ``Heterogeneous-Elasticity Theory'' (HET) that was originally developed for related questions in glass physics (e.g., the boson peak and the scattering of long-wavelength waves by glassy disorder~\cite{Schirmacher_2006,Schirmacher_prl_2007,Schirmacher_2013_boson_peak,schirmacher2015theory}), has been recently applied to nonphononic excitations in glasses~\cite{schirmacher_cutoff_nat_comm_2024}. HET belongs to a class of effective medium theories in which the effect of spatially fluctuating elastic moduli in finite-dimensional continua is treated through various field-theoretic, mean-field approximations. It typically culminates in a set of complex (in the complex variables theory sense), frequency-dependent Green's functions, elastic moduli and elastic wave-speeds, which represent effective, disordered-averaged media~\cite{thorpe_emt_1985,mw_EM_epl,lubensky_emt_2010}.

In~\cite{schirmacher_cutoff_nat_comm_2024}, it is argued that HET --- and a generalization of it discussed therein, denoted as GHET --- give rise to two distinct types of nonphononic, low-frequency excitations, each accompanied by a different nonphononic VDoS ${\cal D}(\omega)$ (denoted as $g(\omega)$ in~\cite{schirmacher_cutoff_nat_comm_2024}). Our goal in this Letter is to quantitatively test the main predictions of~\cite{schirmacher_cutoff_nat_comm_2024} regarding nonphononic excitations as observed in computer glasses. For the most part, we will neither discuss the assumptions and approximations involved in HET/GHET, nor their justification and validity, but rather focus on the emerging testable predictions.

To set the stage for the analysis to follow, we list here the most notable predictions of~\cite{schirmacher_cutoff_nat_comm_2024}:
\begin{enumerate}[leftmargin=*]
    \item[(i)] The nonphononic (nonDebye) VDoS of computer glasses quenched from high parent-temperature liquid states is dominated by `type-I' nonphononic excitations, which follow ${\cal D}(\omega)\!\sim\!\omega^2$, rather than ${\cal D}(\omega)\!\sim\!\omega^4$. Similar predictions have been made in earlier works, e.g.,~\cite{eric_boson_peak_emt,silvio}.
    \item[(ii)] `Type-I' nonphononic excitations correspond to random-matrix-like modes that are spatially extended~\cite{Beltukov2011}.
    \item[(iii)] The nonphononic VDoS of computer glasses quenched from low parent-temperature liquid states (`more stable' in the terminology of~\cite{schirmacher_cutoff_nat_comm_2024}) is dominated by `type-II' nonphononic excitations, which follow a nonuniversal ${\cal D}(\omega)\!\sim\!\omega^{\beta(\alpha)}$. It is argued that the nonuniversal power-law exponent $\beta(\alpha)\=5\!-\!2\alpha$ emerges from a fundamental relation between the nonphononic VDoS and the statistics of the microscopic, interatomic forces/stresses $\sigma$ (cf.~Eq.~(8) in~\cite{schirmacher_cutoff_nat_comm_2024}), which assumes a singular $p(\sigma)\!\sim\!\sigma^{-\alpha}$ distribution at small stresses. Note that $\sigma\=\sigma_{ij}$, where $\sigma_{ij}$ is the force between interacting particles $i$ and $j$.
    \item[(iv)] The exponent $\alpha$ satisfies $\alpha(m)\!=\!1\!-\!1/m$ for purely repulsive interatomic potentials, where $m$ corresponds to the number of derivatives that are continuous at a cutoff distance at which the interatomic interaction in computer glass formers is tailored to vanish (the \emph{tapering function}). Points (iii)-(iv) imply $\beta(m)\!=\!3+2/m$, and it is further argued that computer glasses employing attractive pairwise interactions correspond to $m\!=\!1$, i.e., should feature $\beta\=5$.
    \item[(v)] The so-called 'type-II' nonphononic excitations feature `non-irrotational', vortex-like displacement fields.
\end{enumerate}
As stated above, our goal here is to quantitatively and directly test assertions (i)-(v) through careful computer simulations and based on analyzing relevant data presented in~\cite{schirmacher_cutoff_nat_comm_2024}.
\begin{figure}[ht!]
  \includegraphics[width = 0.49\textwidth]{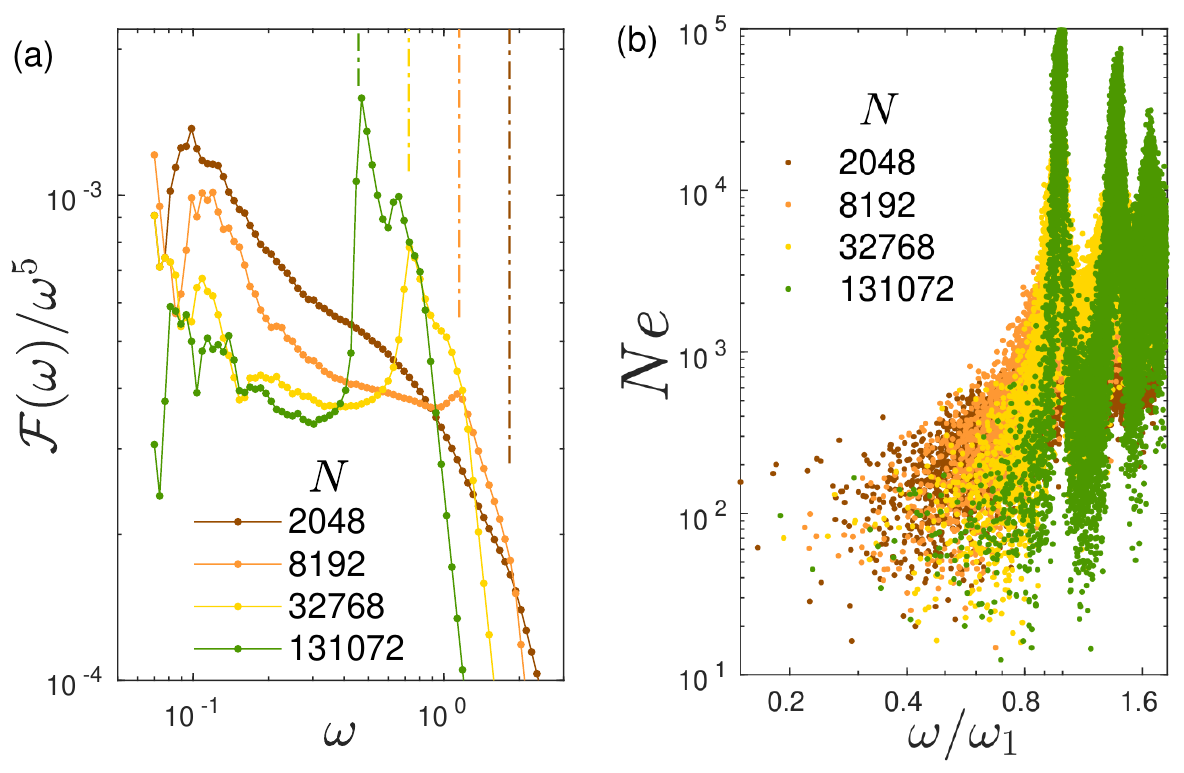}
  \vspace{-0.25cm}
  \caption{\footnotesize Testing claims (i)-(ii) of~\cite{schirmacher_cutoff_nat_comm_2024} regarding `type-I' nonphononic excitations, see list in the text. (a) Rescaled cumulative VDoS ${\cal F}(\omega)/\omega^5$ calculated for IPL glasses of various sizes $N$ (see legend), quenched instantaneously from high-temperature equilibrium states, see text for discussion. The vertical dashed lines mark the lowest phonon frequency $\omega_1\!=\!2\pi c_{\rm s}/L$ for each $N$, where $c_{\rm s}$ denotes the speed of shear waves, and $L$ is the linear system size. Deviations from the universal quartic law (corresponding to ${\cal F}(\omega)\!\sim\!\omega^5$) appear to be larger for smaller $N$. (b) The scaled participation ratio $Ne$ (see text for definition) scatter-plotted against $\omega/\omega_1$. We observe that $Ne\!\sim\!{\cal O}(100)$ as $\omega\!\to\!0$, indicating the (quasi)localization of vibrations.}
  \label{fig:type_I_fig}
\end{figure}

We first address claims (i)-(ii) above. To this aim, we prepared computer glasses quenched from high parent-temperature equilibrium liquid states using a canonical 3D computer glass-former, known as the inverse-power-law (IPL) model, see model details in~\cite{cge_paper}. The system is first equilibrated at a high temperature of $T\!=\!2.0$ (in simulational units~\cite{cge_paper}), which is roughly 4 times larger than the respective computer glass transition temperature, followed by an instantaneous quench to zero temperature by minimizing the potential energy using a standard conjugate gradient algorithm. We then calculated the low-frequency spectra of ensembles of glasses with different system sizes $N$, and measured the participation ratio $e\!\equiv\!\big[N\sum_i(\Psiv_i\cdot\Psiv_i)^2\big]^{-1}$ ($i$ denotes particle indices) of each individual vibrational mode $\Psiv$.

In Fig.~\ref{fig:type_I_fig}a, we present the cumulative VDoS ${\cal F}(\omega)\!\equiv\!\int_0^\omega\!{\cal D}(\omega')\,d\omega'$, rescaled by $\omega^5$, for various $N$'s (see legend). If the nonphononic spectrum follows ${\cal D}(\omega)\!\sim\!\omega^4$, one expects ${\cal F}(\omega)/\omega^5$ to be constant as $\omega\!\to\!0$ below the first phonon band~\cite{phonon_widths,JCP_Perspective} (marked in the figure). It is observed that deviations from the universal quartic law (i.e., from a constant in this presentation) appear to be larger for smaller $N$. These results strongly suggest that claim (i) regarding possible deviations from a universal ${\cal D}(\omega)\!\sim\!\omega^4$ simply reflects finite-size effects, which are unfortunately not discussed in~\cite{schirmacher_cutoff_nat_comm_2024}. In Fig.~\ref{fig:type_I_fig}b, we scatter-plot $Ne(\omega)$ for the glasses used in panel (a). For localized modes, one has $Ne\!\sim\!{\cal O}(100)$ independently of $N$, for $\omega\!\to\!0$~\cite{footnote}, as indeed observed, in contrast to claim (ii) that low-frequency nonphononic modes in glasses quenched from high parent temperatures are spatially extended, in which case $Ne$ should scale as $\sim\!N$.

\begin{figure}[ht!]
  \includegraphics[width = 0.5\textwidth]{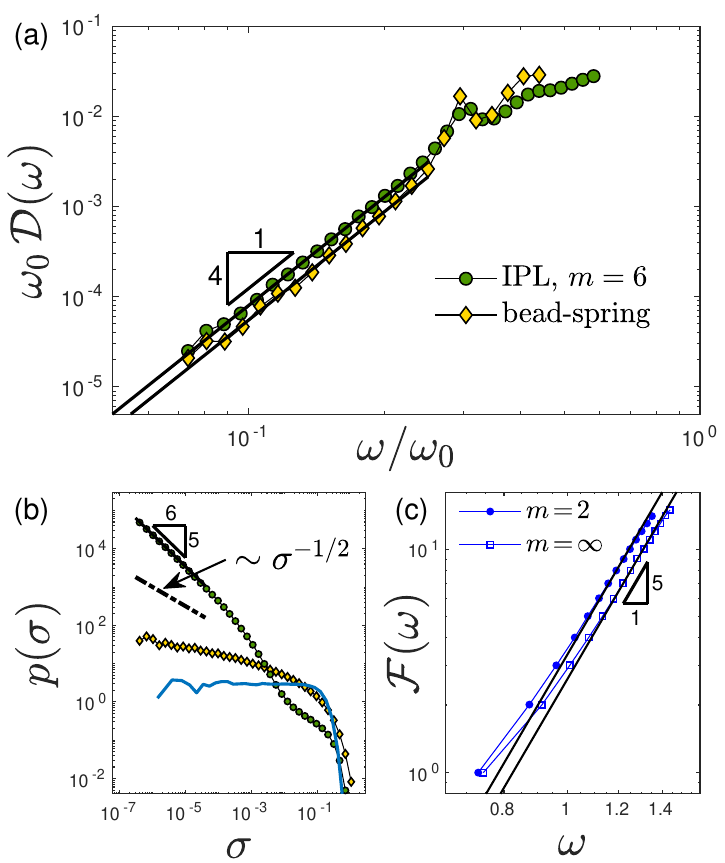}
  \vspace{-0.5cm}
  \caption{\footnotesize Testing claims (iii)-(iv) of~\cite{schirmacher_cutoff_nat_comm_2024} regarding `type-II' nonphononic excitations, see list in the text. (a) ${\cal D}(\omega)$ for the IPL (circles) and bead-spring (diamonds) models (see text), corresponding to $m\!=\!6$ and $m\!=\!1$, respectively. Here, $\omega_0\!\equiv\!c_{\rm s}/a_0$, where $a_0$ is an interatomic distance. According to~\cite{schirmacher_cutoff_nat_comm_2024}, these VDoSs should feature $\beta\!=\!10/3$ for the IPL and $\beta\!=\!5$ for the bead-spring models, respectively. Instead, we find the universal $\beta\!=\!4$ for both. (b) Distributions $p(\sigma)$ of pairwise contributions $\sigma$ to the stress~\cite{footnote4}, measured in the bead-spring and IPL models, see legend of panel (a). The dash-dotted line corresponds to $p(\sigma)\!\sim\!\sigma^{-1/2}$, and the continuous line is the distribution of particlewise stresses $\sigma_i\!\equiv\!\sum_j\sigma_{ij}$ of the same IPL model, see text for discussion.
  (c) Digitized data from Fig.~4b of~\cite{schirmacher_cutoff_nat_comm_2024} for two cumulative VDoSs ${\cal F}(\omega)\!\equiv\!\int_0^\omega\!{\cal D}(\omega')\,d\omega'$, both consistent with ${\cal D}(\omega)\!\sim\!\omega^4$ (see black lines and power-law triangle), see text for discussion.}
  \label{fig:cutoff}
\end{figure}

We next address claims (iii)-(v) in the list above. First, we employ the same IPL model as above (with $N\!\approx\!8000$ particles), but this time generate glassy states corresponding to low parent temperatures (`more stable' glasses) by performing a continuous quench at a rate $\dot{T}\!=\!10^{-3}$ (in simulational units~\cite{cge_paper}) and set $m\!=\!6$ in the tapering function. According to claim (iii), one expects ${\cal D}(\omega)$ to deviate from ${\cal D}(\omega)\!\sim\!\omega^{4}$ at low $\omega$. Specifically, according to claim (iv) one should observe ${\cal D}(\omega)\!\sim\!\omega^{10/3}$, since $m\=6$. In contrast, we show in Fig.~\ref{fig:cutoff}a that the nonphononic VDoS follows the universal quartic scaling, ${\cal D}(\omega)\!\sim\!\omega^4$, in agreement with a large body of existing evidence~\cite{JCP_Perspective}.
Second, we consider a bead-spring model of glass elasticity (see~\cite{frustrated_networks_arXiv_2024} for details) that is regarded as a `stable glass`~\cite{footnote2}, in which there is no interaction cutoff and beads experience both repulsive and attractive forces. According to claim (iv), the latter corresponds to $m\=1$ and hence one expects ${\cal D}(\omega)\!\sim\!\omega^5$. However, we show in Fig.~\ref{fig:cutoff}a that also here ${\cal D}(\omega)\!\sim\!\omega^4$.

\begin{figure*}[ht!]
  \includegraphics[width = 1.0\textwidth]{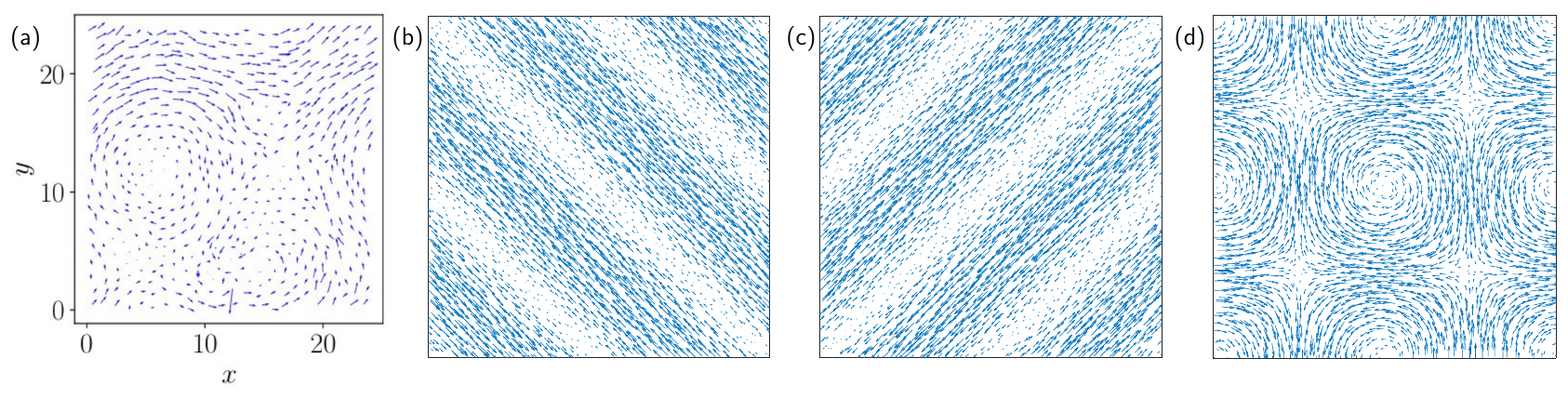}
  \vspace{-0.5cm}\caption{\footnotesize (a) Reproduction of Fig.~1 of the published \href{https://static-content.springer.com/esm/art\%3A10.1038\%2Fs41467-024-46981-7/MediaObjects/41467_2024_46981_MOESM1_ESM.pdf}{Peer Review File} of~\cite{schirmacher_cutoff_nat_comm_2024}, showing a 2D slice of a claimed `Type-II' nonphononic excitation in a computer glass. Plotted is a 2D slice of the 12$^{\mbox{\tiny th}}$ vibrational mode (ordered by frequency), which is almost always a (hybridized) phonon in computer glasses -- since the lowest-frequency phonon-band consists of 12 modes~\cite{phonon_widths}. Panels (b) and (c) show 2D transverse phonons of identical wavelength, but orthogonal wave-vectors. The superposition of these phonons is shown in panel (d). See text for discussion.}
  \label{fig:phonons}
\end{figure*}

Since $\beta\=4$ in both datasets in Fig.~\ref{fig:cutoff}a, the relation $\beta\!=\!5\!-\!2\alpha$ in claim (iii) suggests that weak interatomic stresses $\sigma$ follow a distribution $p(\sigma)\!\sim\!\sigma^{-1/2}$. Here, $\sigma\!=\!a_0^{-3}\partial\varphi/\partial\gamma$, where $\gamma$ is a shear-strain parameter, $\varphi$ is a pairwise potential, and $a_0\!\equiv\!(V/N)^{1/3}$ is a microscopic length with $V$ denoting the system's volume. In Fig.~\ref{fig:cutoff}b, we plot $p(\sigma)$~\cite{footnote4} for the two datasets presented in Fig.~\ref{fig:cutoff}a, revealing that none of them follow $p(\sigma)\!\sim\!\sigma^{-1/2}$. Note that for the IPL model we find $p(\sigma)\!\sim\!\sigma^{-5/6}$ at small $\sigma$ (see power-law triangle in the figure), as anticipated from the relation $\alpha\!=\!1\!-\!1/m$ in claim (iv) with $m\=6$, yet the $\beta\!=\!5\!-\!2\alpha$ prediction in claim (iii) is strongly violated.

While our main goal, as stated above, is to test the predictions in points (i)-(v) listed above, and not to assess the assumptions and approximations that lead to them, one may still wonder how come a coarse-grained, long-wavelength approach such as HET/GHET could possibly invoke the distribution of microscopic, interatomic contributions to the stress $p(\sigma)\!\sim\!\sigma^{-\alpha}$. To shed additional light on this point, beyond the clear violation of the suggested relation $\beta\!=\!5\!-\!2\alpha$ in Fig.~\ref{fig:cutoff}a-b, we also consider next a minimal coarse-graining of the local stress $\sigma$ in order to explore its effect on $p(\sigma)$.

To this aim, we explicitly denote the pairwise contribution to the stress as
$\sigma_{ij}$ (that was so far denoted just as the interatomic stress $\sigma$) and apply a minimal coarse-graining procedure in which we define the particlewise stresses $\sigma_i\!\equiv\!\sum_j\sigma_{ij}$, where particle $j$ is an interacting neighbor of particle $i$~\cite{harrowell_stress_correlations_jcp_2016}. The distribution $p(\sigma_i)$ is plotted for the IPL model (with $m\!=\!6$) with a continuous blue line in Fig.~\ref{fig:cutoff}b. It is observed that the $\sigma_{ij}\!\to\!0$ singularity seen in $p(\sigma_{ij})$ (previously denoted as $p(\sigma)$, i.e., the divergence of $p(\sigma)$ with vanishing pairwise stress, $\sigma\!\to\!0$) due to the tapering of the interaction potential does \emph{not} carry over to the stresses minimally coarse-grained over the particles' scale. This result is generically expected for any glass, independently of any tapering of the interaction potential; it is a simple outcome of the dominance of large contributions to the stress over weak ones, which may be the origin of the failure of claims (iii)-(iv).

In an attempt to support claim (iv), Fig.~4 in~\cite{schirmacher_cutoff_nat_comm_2024} presents results of 3D computer simulations with $N\!=\!1000$ particles; yet, no tests for finite-size effects were reported, despite that it has been repeatedly shown~\cite{lerner2019finite,jcp_2d_vdos_2022,qlm_detection_paper_pre_2023,Ning_2024_finite_size_nat_comm} that finite-size deviations from the $\sim\!\omega^4$ law generally exist, cf.~Fig.~\ref{fig:type_I_fig}a. These finite-size effects can persist up to $N\!\sim\!10^5$~\cite{lerner2019finite}, and are more pronounced in systems cooled quickly from higher equilibrium parent temperatures~\cite{jcp_2d_vdos_2022,Ning_2024_finite_size_nat_comm}. In Fig.~\ref{fig:cutoff}c, we show two datasets (for two $m$ values, see legend) digitized from Fig.~4b of~\cite{schirmacher_cutoff_nat_comm_2024} for the cumulative VDoS ${\cal F}(\omega)$; in~\cite{schirmacher_cutoff_nat_comm_2024}, it is argued that the $m\!=\!\infty$ dataset features $\beta\!=\!3$ (thus ${\cal F}(\omega)\!\sim\!\omega^4$). However, we show here that both the $m\!=\!2$ and $m\!=\!\infty$ dataset are consistent with ${\cal F}(\omega)\!\sim\!\omega^5$ (black lines). The deviations seen in the lowest frequencies are known finite-size effects~\cite{qlm_detection_paper_pre_2023,Ning_2024_finite_size_nat_comm}, as also shown in Fig.~\ref{fig:type_I_fig}a.

In~\cite{schirmacher_cutoff_nat_comm_2024}, the `non-irrotational', vortex-like real-space structure of `type-II' nonphononic excitations, i.e., claim (v) above, is extensively discussed. Yet --- and unfortunately --- the actual spatial structures are not presented therein. However, an example of a non-irrotational nonphononic excitation is presented in Fig.~1 of the published \href{https://static-content.springer.com/esm/art\%3A10.1038\%2Fs41467-024-46981-7/MediaObjects/41467_2024_46981_MOESM1_ESM.pdf}{Peer Review File} of~\cite{schirmacher_cutoff_nat_comm_2024}, which is reproduced here in Fig.~\ref{fig:phonons}a. Plotted is a 2D slice of the 12$^{\mbox{\tiny th}}$ vibrational mode (being ordered by frequency), which features vortex-like patterns. Such vortex-like displacement fields generally emerge from a superposition of {\em phonons} of similar frequencies, as demonstrated in Fig.~\ref{fig:phonons}b-d. Indeed, the 12$^{\mbox{\tiny th}}$ vibrational mode is almost always a (hybridized) phonon in computer glasses -- since the lowest-frequency phonon-band consists of 12 modes~\cite{phonon_widths}. The same vortex-like patterns seen in Fig.~\ref{fig:phonons}a are observed in vibrational excitations up to and in the vicinity of the boson-peak frequency, as demonstrated recently in~\cite{boson_peak_2d_jcp_2023}.

\begin{figure*}[ht!]
  \includegraphics[width = 1.0\textwidth]{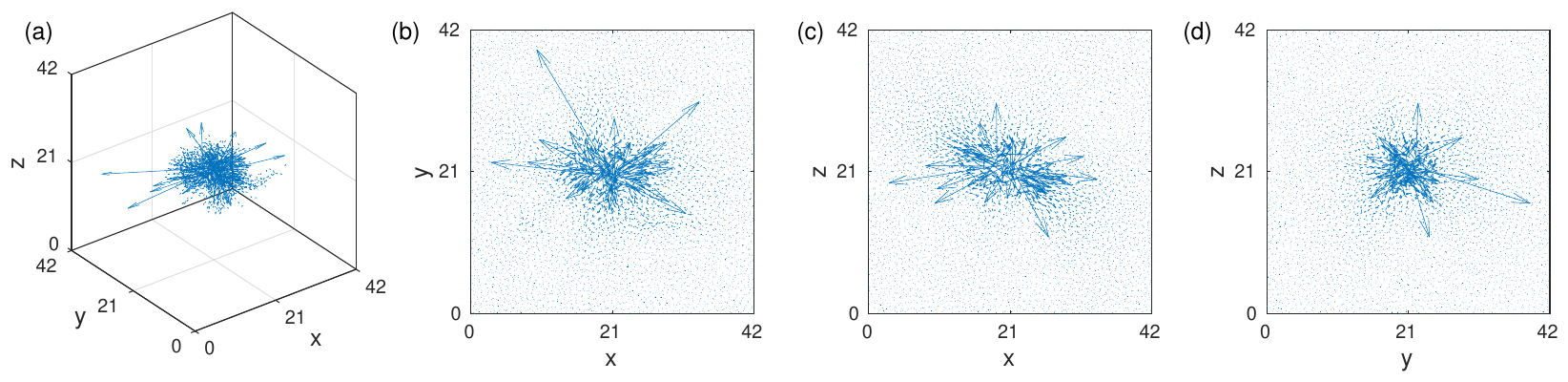}
  \vspace{-0.5cm}\caption{\footnotesize (a) A typical low-frequency quasilocalized vibrational mode measured in a slowly quenched IPL computer glass of $N\!=\!64000$ particles. Only the 2\% largest components are shown. This mode has a frequency $\omega/\omega_1\!=\!0.26$, i.e.,~it dwells far below the lowest frequency phonons. (b)-(d) show projections of thin slices of the mode onto all 3 Cartesian planes; here, no filtering by components' magnitude was applied. We see no signs of vorticity in the mode's spatial structure, at odds with the claims of~\cite{schirmacher_cutoff_nat_comm_2024} that nonphononic modes feature non-irrotational fields, and with the mode shown in Fig.~\ref{fig:phonons}a.}
  \label{fig:qlm}
\end{figure*}

In Fig.~\ref{fig:qlm}, we display a typical low-frequency quasilocalized mode that features a frequency $\omega\!\sim\!\omega_1/4$, i.e.~well below the lowest-frequency phonons. It shows no signs of vorticity in its spatial structure. Consequently, we suggest that the object claimed in~\cite{schirmacher_cutoff_nat_comm_2024} to be a `non-irrotational' {\em nonphononic} excitation is, in fact, of {\em phononic} nature and hence that claim (v) involves a misinterpretation.

To conclude, we showed that the nonphononic VDoS of computer glasses follows the universal $\sim\!\omega^4$ quartic law, independently of the form of the tapering function of the pairwise potential, of the glass formation history, and of the form of the distribution of pairwise contributions to the stress -- all at odds with the claims of~\cite{schirmacher_cutoff_nat_comm_2024}. We also showed that nonphononic excitations in glasses quenched from high parent temperatures are (quasi)localized~\cite{footnote}, that minimally coarse-graining the interatomic stresses --- even on the smallest possible scale --- removes the potential-tapering-induced singularity of the atomistic stress distribution, and that the vortex-like patterns seen in low-frequency modes naturally correspond to a superpositions of phonons of similar frequencies, again at odds with the claims of~\cite{schirmacher_cutoff_nat_comm_2024}.

Our findings thus highlight the need for additional theoretical developments in relation to low-frequency, nonphononic excitations in glasses. Moreover, they suggest that theories --- such as the Heterogeneous-Elasticity Theory --- that are formulated in terms of coarse-grained, continuum fields are likely to be inherently deficient in capturing the intrinsically atomistic and micromechanical nature of quasilocalized, nonphononic excitations in structural glasses.

\emph{Acknowledgements}.--- We thank Francesco Zamponi and Walter Schirmacher for open and useful discussions. This work has been supported by the Israel Science
Foundation (ISF grant no.~403/24). E.B.~is supported by the Ben May Center for Chemical Theory and Computation, and the Harold Perlman Family.


\begin{thebibliography}{61}
\bibitem [{foo({\natexlab{c}})}]{footnote4}%
  \BibitemOpen
  \bibinfo {note} {We rescale the pairwise contributions to the stress by the
  shear modulus for both studied models, to the aim of comparing them on equal
  footing.}\BibitemShut {Stop}%
\end{thebibliography}
%

\end{document}